\documentclass{aa}
\usepackage{graphicx}
\input{psfig.sty}
\newcommand{\beq}{\begin{equation}}
\newcommand{\eeq}{\end{equation}}
\newcommand{\ba}{\begin{array}}
\newcommand{\ea}{\end{array}}

\begin{document}

\title{ Jeans' gravitational instability and  nonextensive kinetic theory}
\author{J. A. S. Lima, R. Silva, J. Santos }

\pretolerance=4000

\institute{Departamento de F\'{\i}sica, UFRN, C.P. 1641\\
           59072-970 Natal, RN, Brazil }

\offprints{Lima (limajas@dfte.ufrn.br)}

\titlerunning{Jeans' gravitational instability and nonextensive kinetic theory}

\date{Received 18 March 2002 /Accepted 27 August 2002}

\abstract{The concept of Jeans gravitational instability is
rediscussed in the framework of nonextensive statistics and its
associated kinetic theory. A simple analytical formula
generalizing the Jeans criterion is derived by assuming that the
unperturbed self-gravitating collisionless gas is kinetically
described by the $q$-parameterized class of power law velocity
distributions. It is found that the critical values of wavelength
and mass depend explicitly on the nonextensive $q$-parameter. The
standard Jeans wavelength derived for a Maxwellian distribution is
recovered in the limiting case $q$=1. For power-law distributions
with cutoff, the instability condition is weakened with the system
becoming unstable even for wavelengths of the disturbance smaller
than the standard Jeans length $\lambda_J$.}

\maketitle

\section{Introduction}

In a seminal paper, James Jeans discussed the conditions under
which a fluid becomes gravitationally unstable under the action of
its own gravity (Jeans 1929). Nowadays, for any relevant length
scales in the Universe (stars, galaxies, clusters, etc), such an
instability has been recognized as the key mechanism to explain
the gravitational formation of structures and their evolution in
the linear regime.

A noteworthy conclusion of Jeans' work is that perturbations with
mass greater than a critical value $M_J$ (Jeans' mass) may grow
thereby producing gravitationally bounded structures, whereas
perturbations with a mass smaller than $M_J$ do not grow and
behave like acoustic waves. This gravitational instability
criterion remains basically valid and plays a fundamental role
even in the expanding Universe (Peebles 1993; Coles \& Lucchin
1995). In terms of the wavelengths of a fluctuation, Jeans'
criterion says that $\lambda$ should be greater than a critical
value $\lambda_J=\sqrt{\pi v_s^2/(G\rho_0)}$, which is named the
{\em Jeans' length}. In this formula $G$ is the gravitational
constant, $\rho_0$ is the unperturbed matter density and $v_s$ is
the sound speed for adiabatic perturbations. As is widely known,
the same criterion also holds for a collisionless self-gravitating
cloud of gas, except that the sound speed $v_s$ is replaced by the
velocity dispersion $\sigma$. As explained in many textbooks (see,
for instance, Binney \& Tremaine 1987), such a result follows
naturally from a kinetic theoretical approach based on the the
Vlasov equation, where the evolution of the collisionless gas is
ultimately described by perturbing the equilibrium Maxwellian
velocity distribution.

On the other hand, some recent studies involving the statistical
description of a large variety of physical systems revealed that
some extension of the standard Boltzmann-Gibbs approach should be
needed. Particular examples are systems endowed with long duration
memory, anomalous diffusion, turbulence in pure-electron plasma,
self-gravitating systems or more generally systems endowed with
long range interactions. In order to work with such problems,
Tsallis (1988) proposed the following generalization of the
Boltzmann-Gibbs (BG) entropy formula for statistical equilibrium
(see also http.tsallis.cat.cbpf.br/biblio.htm for an extensive and
updated list of problems and references)
\begin{equation} S_q = -k_B\left(1-\sum_{i}
p_i^{q}\right)/(q-1)\quad, \end{equation} where $k_B$ is the
Boltzmann constant, $p_i$ is the probability of the $i$th
microstate and $q$ is a parameter quantifying the degree of
nonextensivity. In the limit $q\rightarrow 1$ the celebrated BG
extensive formula, namely
\begin{equation} S=-k_B\sum_{i} p_i\ln p_i\quad, \end{equation}
is recovered. One of the main properties of $S_q$ is its
pseudoadditivity. Given a composite system $A+B$, constituted by
two subsystems $A$ and $B$, which are independent in the sense of
factorizability of the microstate probabilities, the Tsallis
measure verifies
$S_q(A+B)=S_q(A)+S_q(B)+(1-q)k_{B}^{-1}S_q(A)S_q(B)$. In the limit
$q\rightarrow 1$, $S_q$ reduces to the standard logarithmic
measure, and the usual additivity of the extensive BG statistical
mechanics and thermodynamics is also recovered. Thus, if $q$
differs from unity, the entropy becomes nonextensive
(superextensive if $q < 1$, and subextensive if $q > 1$), with the
Boltzmann factor generalized into a power law. In other words,
$|q-1|$ quantifies the lack of extensivity of the system.

A large portion of the experimental evidence supporting Tsallis
proposal involves a non-Maxwellian (power-law) equilibrium
distribution function associated with the thermostatistical
description of the classical $N$-body problem. This equilibrium
velocity $q$-distribution may be derived from at least three
different methods: (i) Through a simple nonextensive
generalization of the Maxwell ansatz, which is based on the
isotropy of the velocity space (Silva et al 1998); (ii) Within the
nonextensive canonical ensemble, that is, maximizing Tsallis
entropy under the constraints imposed by normalization and the
energy mean value (Abe 1999) (iii) Using a more rigorous treatment
based on the nonextensive formulation of the Boltzmann $H$-theorem
(Lima et al. 2001).

In the astrophysical context, the nonextensive equilibrium
velocity distribution related to Tsallis' statistics has been
applied to stellar collisionless systems (Plastino \& Plastino
1993), as well as to the peculiar velocity function of galaxies
clusters (Lavagno 1998). In the former paper, the entropy
nonextensive formula was maximized taking into account the
constraints imposed by the total mass and energy density. It was
found that the solutions behave like stellar polytropes and that
the mass is finite if the nonextensive parameter is bounded by $q
\leq 9/7$. More recently, a more detailed account for spherically
symmetric systems involving either the gravothermal instability
(Taruya \& Sakagami 2002, Chavanis 2002), or the complete
determination of the density profiles has been presented in the
literature (Lima \& de Souza 2002).

Applications in plasma physics are also important in their own
right, both from a methodological and physical viewpoint. In this
case, some metastable states in pure electron plasmas, and the
dispersion relations for an electrostatic plane-wave propagation
in a collisionless thermal plasma (including undamped Bohm-Gross
and Landau damped waves) have also been studied and compared to
the standard results (Boghosian 1996; Lima et al 2000). In
particular, Liu et al (1994) showed a reasonable indication for
the non-Maxwellian velocity distribution from plasma experiments.
All this empirical evidence deal directly or indirectly, with the
$q$-distribution of velocities for a nonrelativistic gas.

In the present work, we quantify to what extent the nonextensive
effects modify the gravitational instability criterion established
by Jeans. For this enlarged framework, we deduce a new analytical
expression for the Jeans dispersion relation which gives rise to
an extended instability criterion in accordance with Tsallis
thermostatistics.

\section{Nonextensive Gravitational Instability}

In what follows, we focus our attention on the kinetic description
of a many particle system within the nonrelativistic gravitational
context. More precisely, we consider an infinite self-gravitating
collisionless gas described by a distribution function $f({\bf
v},{\bf r},t)$ slightly depart from equilibrium. If $f_0(v)$
corresponds to the unperturbed homogeneous and time-independent
equilibrium distribution, sometimes referred to as the ``Jeans
Swindle", the resulting particle distribution function may be
approximated as
\begin{equation} \label{fundist}
f = f_0(v) + f_1({\bf v}, {\bf r}, t)\quad,\quad f_1\ll f_0\quad,
\end{equation}
where $f_1$ is the corresponding perturbation in the distribution
function. Following standard lines, the dynamic behavior of this
system can be described by the linearized Vlasov and Poisson
equations. By neglecting up to second-order terms in the expansion
of the distribution function one obtains

\begin{equation}
\label{Boltz} \frac{\partial f_1}{\partial t} + {\bf
v}.\frac{\partial f_1}{\partial {\bf r}} =
{\nabla}\phi_0.\frac{\partial f_1}{\partial {\bf v}} +
{\nabla}\phi_1.\frac{\partial f_0}{\partial {\bf v}}\quad,
\end{equation}

\begin{equation}
{\nabla^2}\phi_1 = 4\pi G{\int f_1({\bf r},{\bf v},t)d^3v}\quad,
\end{equation}
where $\phi_0({\bf r})$ and $\phi_1({\bf r})$ are, respectively,
the unperturbed gravitational potential and its first order
correction. As is well known, if the ``Jeans Swindle" is assumed,
one may set the unperturbed potential $\phi_0=0$. In this case,
the solutions of the above equations can be written as $f_1({\bf
r},{\bf v},t)=F({\bf v})\exp\{i({\bf k}\cdot{\bf r}-\omega t)\}$
and $\phi_1({\bf r},t)=\phi\exp\{i({\bf k}\cdot{\bf r}-\omega
t)\}$ provided that $F({\bf v})$ and $\phi$ satisfy the relations
\begin{equation}
\label{relations} ({\bf k}\cdot{\bf v} - \omega)F({\bf v}) -
\phi{\bf k}.\frac{\partial f_0}{\partial{\bf v}} = 0\quad,
\end{equation}
\begin{equation}
k^2\phi = - 4\pi G\int F({\bf v})d^3v\quad.
\end{equation}
Combining these expressions one obtains the dispersion relation
between $\omega$ and ${\bf k}$
\begin{equation}
\label{dispersion} 1 + \frac{4\pi G}{k^2}\int\frac{{\bf
k}\cdot\,\partial f_0/\partial{\bf v}}{{\bf k}\cdot{\bf v}-\omega}
d^3v = 0\quad.
\end{equation}
The standard instability analysis follows from the above
dispersion relation when one takes $f_0({\bf v})$ as the
Maxwellian velocity distribution.

Consider now the $q$-nonextensive framework proposed by Tsallis.
In this case, the equilibrium distribution function $f_0({\bf v})$
can be written as (Silva et al 1998; Lima et al 2000)
\begin{equation} \label{f-q} f_0({\bf
v})=\frac{\rho_0B_q}{(2\pi\sigma^2)^{3/2}}
\left[1-(q-1)\frac{v^2}{2\sigma^2}\right]^{1 \over{q-1}} \quad,
\end{equation}
where the normalization constant reads
\begin{equation}  \label{Aq}
B_{q} = \frac{(3q-1)(q+1)}{4}\; \frac{\sqrt{q-1}\;\Gamma({1\over
q-1}+{1\over 2})}{ \Gamma({1\over q-1})}
\end{equation}
for $q\geq 1$, and
\begin{equation} \label{Aq_}
B_{q} =
\left(\frac{3q-1}{2}\right)\frac{\sqrt{1-q}\;\Gamma\left(\frac{1}{1-
q}\right)}{\Gamma\left(\frac{1}{1-q}-\frac{1}{2}\right)}
\end{equation}
for ${1\over 3}<q<1$. Here $\sigma$ is the velocity dispersion and
$\rho_0$ is the equilibrium density. As one may check, for
$q<1/3$, the $q$-distribution (\ref{f-q}) is unnormalizable, while
for $q>1$, it exhibits a thermal cutoff on the maximum value
allowed for the velocity of the particles, which is given by
\begin{equation}
\label{vm} v_{max}=\sqrt{2\sigma^2\over q-1}\quad.
\end{equation}
This thermal cut-off is absent when $q \leq 1$, that is, $v_{max}$
is also unbounded for these values of the $q$-parameter. In this
connection, it is worth noticing that the spirit of the
$H$-theorem is totally preserved for this nonextensive velocity
distribution. As a matter of fact, by introducing a generalized
collisional term, $C_q(f)$, it has been shown that the entropy
source is definite positive for $q > 0$, and does not vanish
unless the $q$-equilibrium distribution function assumes the above
power-law form (Lima et al. 2001). In addition, taking into
account that $\lim_{|z|\rightarrow
\infty}\Gamma(z+a)/[z^a\Gamma(z)]=1$, as well as that
$\lim_{q\rightarrow 1}[1-(q-1)x^2]^{1/(q-1)}=\exp(-x^2)$
(Abramowitz \& Stegun 1972), the expressions (\ref{Aq}) and
(\ref{Aq_}) defining $B_q$ reduce to $B_{1}=1$, and as should be
expected the distribution function $f_0$ reduces to the
exponential Maxwellian distribution
\begin{equation} \label{f-maxwell} f_0({\bf
v})=\frac{\rho_0}{(2\pi\sigma^2)^{3/2}}\exp(-\frac{v^2}{2\sigma^2})\quad.
\end{equation}
In Fig. 1 we have plotted the nonextensive distribution as a
function of ${v}$ for some selected values of the $q$ parameter.
As can be seen, for $q > 1$ the distribution exhibits a cutoff on
the maximum value allowed for the velocity of the particles. In
the limit $q\rightarrow 1$ we see from (\ref{vm}) that
$v_{max}\rightarrow \infty$ and the power-law behavior reduces to
the exponential Maxwellian distribution. For $q < 1$ the cutoff is
also absent with the curves decreasing less rapidly than in the
Maxwellian case. Note also the convergence from power-law to the
standard Gaussian curve as long as the $q$ parameter approaches
unity.

\begin{figure}
\vspace{.2in}
\centerline{\psfig{figure=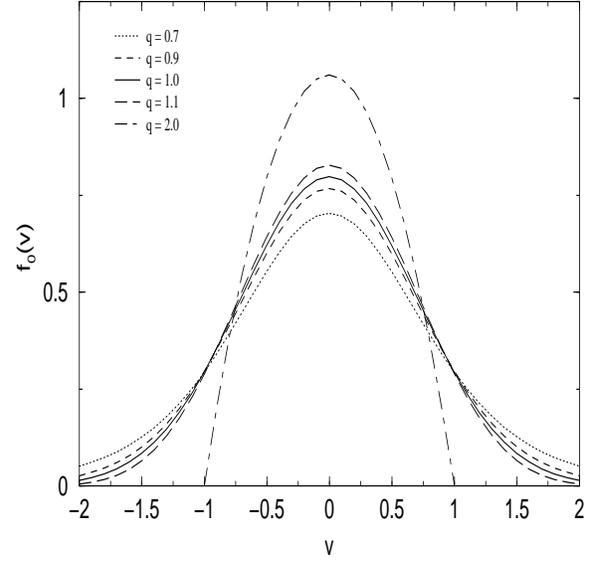,width=3.0truein,height=3.0truein}
\hskip 0.1in} \caption{Nonextensive velocity distribution function
$f_0({\bf v})$ for typical values of the $q$ parameter. Power law
distributions with $q>1$ exhibits a thermal cutoff on the maximum
value allowed for the velocity of the particles. This cutoff is
absent for ${1 \over 3} < q \leq 1$, and the distribution is
unnormalizable for $q < {1 \over 3}$.}
\end{figure}

Now, substituting the nonextensive equilibrium distribution given
by (\ref{f-q}) into the dispersion relation (\ref{dispersion}) one
finds
\begin{equation} \label{dispersion-2} 1-\frac{4\pi G\rho_0}{\sigma^5}
\frac{B_q}{(2\pi)^{3/2}}\int \frac{v_x
\left[1-(q-1)\frac{v^2}{2\sigma^2} \right]^{\frac{2-q}{q-1}}}{kv_x
- \omega} d^3v = 0
\end{equation}
where the $v_x$ axis has been chosen in the direction of ${\bf
k}$.

In the study of Jeans instability, the boundary between stable and
unstable solutions is achieved by setting $\omega =0$ in
(\ref{dispersion-2}). For this value of $\omega$ the above
integral can be easily evaluated and the result is a
$q$-parameterized family of critical wavenumbers $k_q$ given by
\begin{equation} \label{k-q}
{k_{q}}^2={k_{J}}^2{(3q - 1)\over 2}
\end{equation}
where ${k_{J}}^2=4\pi G\rho_0/\sigma^2$ is the classical {\em
Jeans wavenumber\/} (Binney \& Tremaine 1987). From (\ref{k-q}) we
have the $q$-wavelength
$\lambda_q=2\pi/k_q=\lambda_J\sqrt{2/(3q-1)}$ where $\lambda_J$ is
the {\em Jeans wavelength}. Note that the classical value of the
Jeans wavenumber and wavelength as obtained from fluid theory are
recovered only if $q=1$. For $q>1$ ($1/3<q<1$) the $q$-wavelength
$\lambda_q$ is decreased (increased) by the factor
$\sqrt{2/(3q-1)}$. Associated with $\lambda_q$ there is a critical
mass $M_q$, defined as the mass contained within a sphere of
diameter $\lambda_q$, which is given by
$M_q=\frac{4\pi}{3}\rho_0(\lambda_q/2)^3=M_J[2/(3q-1)]^{3/2}$,
where $M_J$ is the critical {\em Jeans mass}. Similarly, we see
that for $q>1$ ($1/3<q<1$) the critical mass is decreased
(increased) by the factor $[2/(3q-1)]^{3/2}$. Therefore, in
comparison with a Maxwellian gas ($q=1$), a nonextensive
collisionless self-gravitating system is more (or less) unstable
depending on the value assumed by the $q$-parameter.

Let us now discuss in more detail the unstable modes. As happens
in the perturbed fluid theory, fluctuations with wavelengths
$\lambda > \lambda_J$ will be unstable. In order to check what
happens in the present kinetic approach we set $\omega=i\gamma$,
where $\gamma$ is real positive, and insert this into the
dispersion relation (\ref{dispersion-2}). As before, choosing the
$v_x$ axis in the direction of ${\bf k}$, the following dispersion
relation is obtained \begin{equation} \label{dispersion-3}
{k^2\over {k_{q}}^{2}}=1-\sqrt{\pi}\beta^2I_q(\beta)\;,
\end{equation} where $\beta =\gamma/\sqrt{2}k\sigma$ and
$I_q(\beta)$ is a $q$-dependent integral given by
\begin{eqnarray*}\label{q1-integral}
I_q(\beta)=\frac{\sqrt{q-1}\Gamma(\frac{1}{q-1}+\frac{1}{2})}
{\Gamma(\frac{1}{q-1})}\frac{q+1}{\pi}
\end{eqnarray*}
\begin{equation}\label{q1-integral}
\times\int_0^{\frac{1}{\sqrt{q-1}}}
\frac{\left[1-(q-1)x^2\right]^{\frac{1}{q-1}}}{\beta^2 + x^2}dx
\end{equation} for $q>1$ and
\begin{equation}\label{q2-integral}
I_q(\beta)=\frac{\sqrt{1-q}\Gamma(\frac{1}{1-q})}
{\Gamma(\frac{1}{1-q}-\frac{1}{2})} \frac{2}{\pi} \int_0^{\infty}
\frac{\left[1-(q-1)x^2\right]^{\frac{1}{q-1}}}{\beta^2 + x^2}dx
\end{equation}
for $1/3<q<1$. It should be noticed that the limit $q\rightarrow
1$ of both expressions is
\begin{equation}
\lim_{q\rightarrow 1}I_q(\beta)=\frac{2}{\pi}\int_0^{\infty}
\frac{e^{-x^2}}{\beta^2 + x^2}dx = \frac{e^{\beta^2}}{\beta}[1 -
{\bf erf}(\beta)]\quad,
\end{equation}
with the expression (\ref{dispersion-3}) reducing to
\begin{equation}
\frac{k^2}{{k_{J}}^2} = 1-{\sqrt{\pi}}\beta e^{\beta^2}[1 - {\bf
erf}(\beta)] \quad,
\end{equation}
which is the standard dispersion relation for stellar systems
(Binney \& Tremaine 1987). For an arbitrary value of $q$, the
dispersion relation is given explicitly by the integral form
(\ref{dispersion-3}) which can be numerically evaluated. In Fig. 2
we plot the unstable branches for several values of $q$ along with
the dispersion relation for an infinite homogeneous fluid which is
given by the solid straight line (see Binney \& Tremaine 1987 for
comparison).

\begin{figure}
\vspace{.2in}
\centerline{\psfig{figure=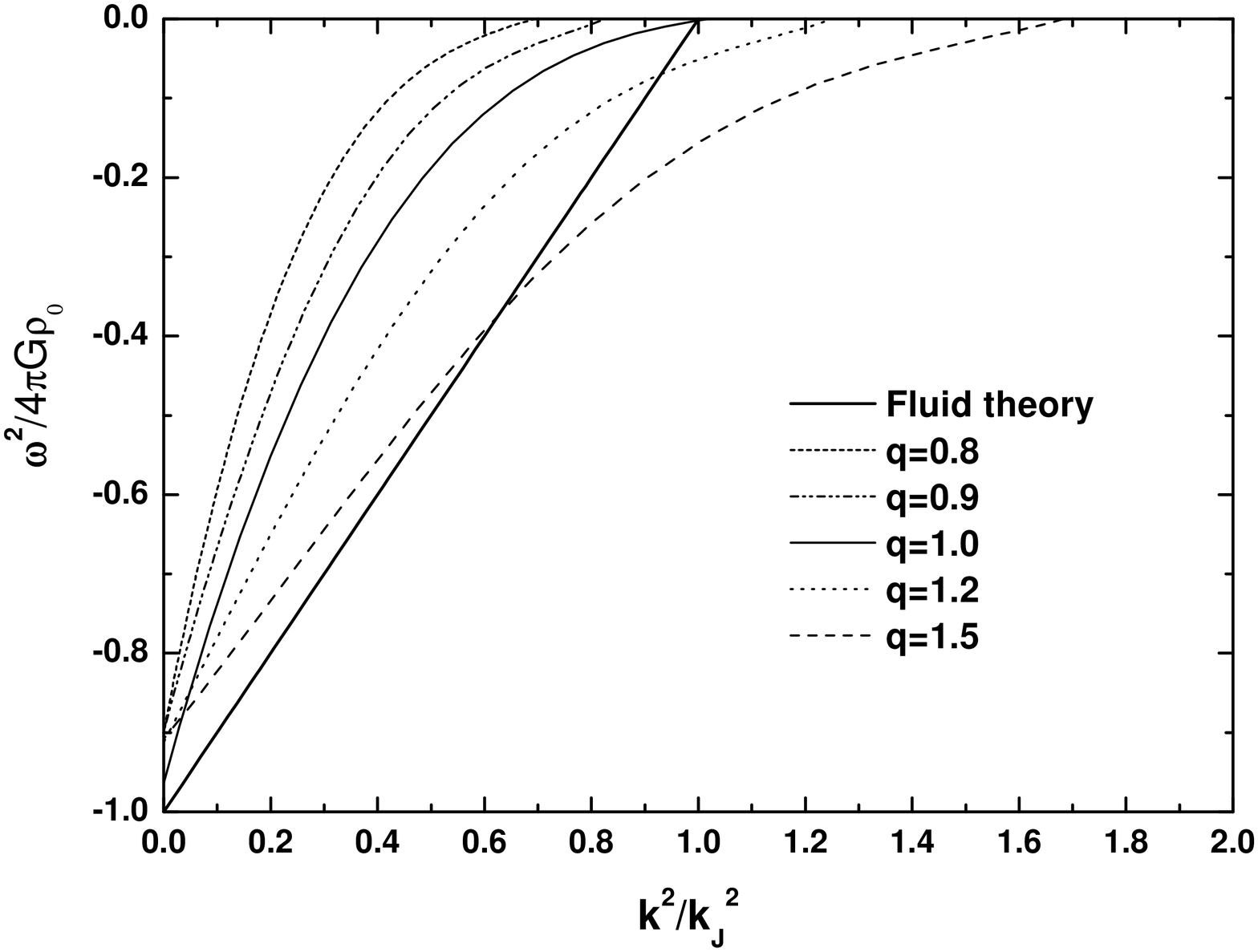,width=3.5truein,height=3.8truein}
\hskip 0.1in} \caption{Unstable branches of the dispersion
relations for an infinite self-gravitating collisionless gas
obeying the nonextensive Tsallis $q$-statistics. For comparison
the straight line representing the dispersion relation for an
infinite homogeneous fluid has also been plotted. We see that for
$q>1$ (power-law distributions with cutoff), the system is
unstable even for wavelengths smaller than the standard Jeans
length value ($\lambda_J$).} \label{}
\end{figure}

As can be seen from these plots, for $q>1$ the system presents
instability even for wavenumbers of the disturbance greater than
the standard critical Jeans value ($k_J$). In other words, this
kind of $q$-gaseous system is more unstable than the one described
by the standard Maxwellian curve, and therefore, structures are
more easily formed in this nonextensive context. However, for
$q<1$ the system may remain stable even for disturbance with a
wavenumber smaller than the Jeans wavenumber. This yields a
generalization of the Jeans gravitational instability mechanism
within Tsallis' nonextensive thermostatistical context. The
standard Jeans criterion is thus modified with critical values of
mass and length depending explicitly on the nonextensive
$q$-parameter.

In Fig. 3 new plots of the dispersion relation (unstable branches)
are presented. Different from Fig.2, instead of the ratio
involving the critical Jeans wavenumber ($k_J$), the behavior of
the unstable branches are analyzed as a function of the ratio
$k/k_q$.

As should be expected at first sight, the curves are convergent
for the critical ratio ($k/k_{q} = 1$), but the growth rates are
slightly different for each value of $q$. Note also that for all
values of $q$, the results are quite different to what happens in
the macroscopic fluid approach for gravitational instability
(solid straight line).

\begin{figure}
\vspace{.2in}
\centerline{\psfig{figure=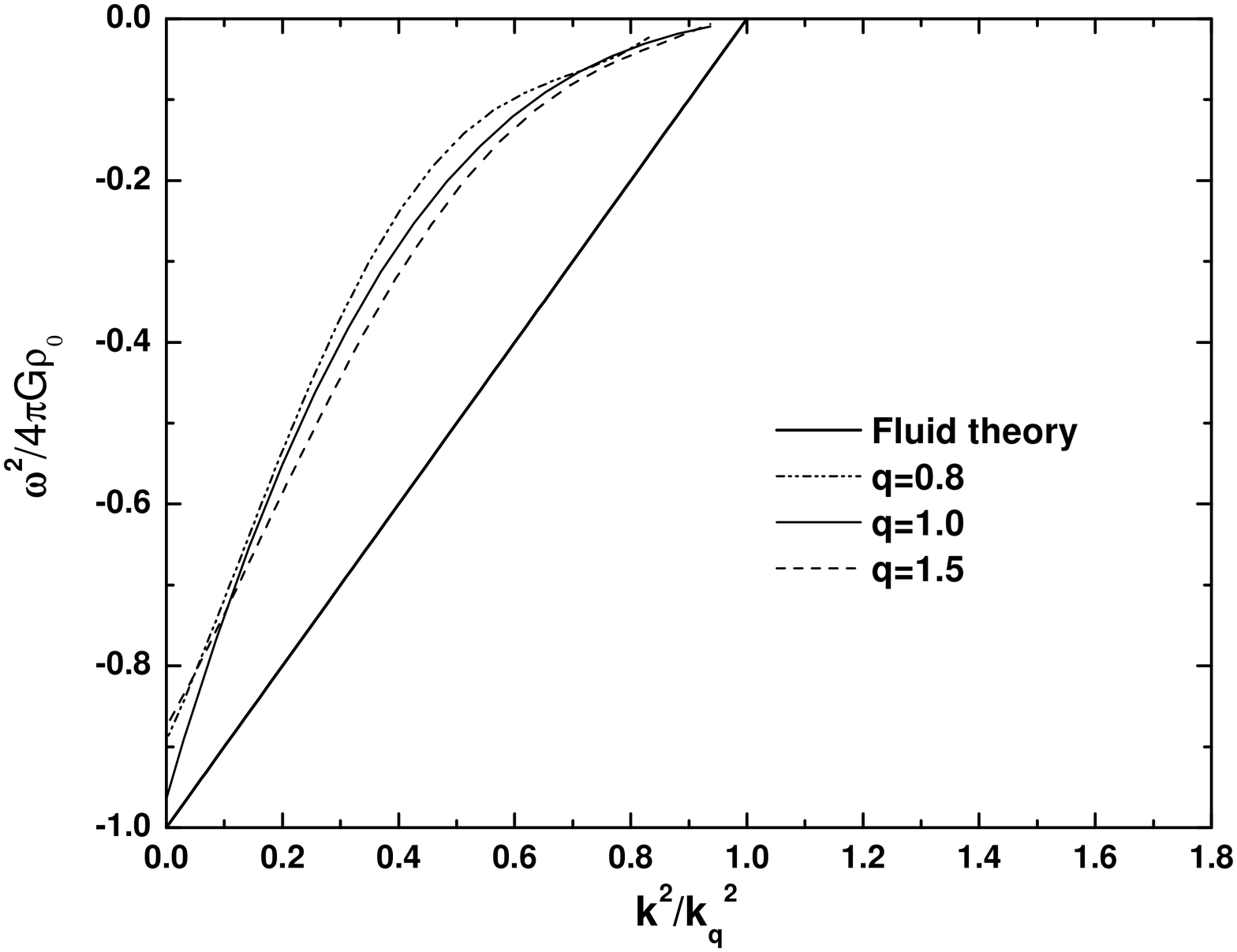,width=3.5truein,height=3.8truein}
\hskip 0.1in} \caption{Unstable branches of the dispersion
relations as a function of the critical wavenumbers $k_q$. Notice
that the magnitude of $k_q$ is related to the critical Jeans
wavenumber by equation \protect(\ref{k-q}). As in Fig.2, the
straight solid line stands to the case of an infinite homogeneous
fluid. }
\end{figure}

\section{Final Remarks}

We have discussed the Jeans gravitational instability along the
lines of the nonextensive statistical formalism proposed by
Tsallis. In this extended kinetic framework, the Gaussian phase
space density is replaced by a family of power law distributions
parameterized by the nonextensive $q$-parameter. An attractive
feature of Tsallis thermostatistics is that the models are
analytically tractable in such a way that a detailed comparison
with the extensive Maxwell-Boltzmann approach is readily achieved.

The main interest of our results rests on the fact that the
Maxwellian curve may provide only a very crude description of the
velocity distribution for a self-gravitating gas, or more
generally for any system endowed with long range interactions. As
we have seen, a well determined criterion for gravitational
instability is not a privilege of the exponential velocity
distribution function, but is shared by an entire family of
power-law functions (named $q$-exponentials) which includes the
standard Jeans result for the Maxwellian distribution as a
limiting case ($q=1$). In general, the basic instability criterion
is maintained: perturbations with $k > k_q$ do not grow (or are
damped even considering that the collisionless $q$-gas is a time
reversible system), while instability takes place for $k < k_q$.
However, unlike Jeans' treatment, in this context there exists a
family of growth rates parameterized by the nonextensive
$q$-parameter. In particular, for $q>1$ (power-law distributions
with cutoff) the system presents instability even for wavenumbers
of the disturbance greater than the standard critical Jeans value
$k_J$.

Finally, since the gravitational instability Jeans' criterion
remains basically true for an expanding Universe (Coles \& Lucchin
1995), the adoption of a power law distribution and the resulting
nonextensive effects may have interesting consequences for the
galaxy formation process. Such implications will be examined in a
forthcoming communication.

\vspace{0.3cm} {\bf Acknowledgments:} The authors are grateful to
Dr. I. Szapudi for many valuable comments. This work was partially
supported by the project Pronex/FINEP (No. 41.96.0908.00), FAPESP
(00/06695-0), and CNPq (62.0053/01-1-PADCT III/Milenio).

\end{document}